\documentstyle{article}
\begin{document}
\begin{center}
\Large\bf{Strange Antibaryon Production Data and Reduction of Strangeness 
Suppression in Sulphur-Nucleus Collisions at 200A GeV}
 \\
\vspace{1cm}\large
 Sa Ben-Hao$^{1,2,3,5}$ and  Tai An$^4$\\ 
\begin{tabbing}
ttttt \= tt \= \kill
\>1. CCAST (World Lab.), P. O. Box 8730 Beijing, China. \\
\>2. China Institute of Atomic Energy, P. O. Box 275 (18), \\
\> \>Beijing, 102413 China.\footnotemark \\
\>3. Department of Physics, University of Hiroshima, \\
\> \>Higashi-Hiroshima, 739 Japan.\\
\>4. Institute of High Energy Physics, Academia Sinica, \\
\> \> P. O. Box 918, Beijing, 100039 China.\\       
\>5. Institute of Theoretical Physics, Academia Sinica, \\
\> \>Beijing China.\\
\end{tabbing}
\end{center}
\footnotetext{mailing address.\\Email: sabh@mipsa.ciae.ac.cn}
\normalsize
\begin{abstract}
We have used the event generator LUCIAE to analyse NA35 data of the $\bar p$ 
and $\bar{\Lambda}$ yields, the ratio $\bar{\Lambda}$/$\bar p$, and the 
transverse mass distributions of $\bar p$ and $\bar{\Lambda}$ in pp and 
central
sulphur-nucleus collisions at 200A GeV. The NA35 data could be reproduced 
reasonably if one assumes that the s quark pair suppression factor and 
concerns in nucleus-nucleus collisions are larger than  the nucleon-nucleon 
collision. It seems to indicate that NA35 data might imply the reduction    
of strangeness suppression in ultrarelativistic nucleus-nucleus collisions 
comparing to the nucleon-nucleon collision at the same energy.  
However, the ratio $\bar{\Lambda}$/$\bar p$ approaching to unity 
in AA collisions comparing to the pp collision does not necessarily mean a flavor 
symmetry since hadronic rescattering plays a role as well.

\end{abstract}
\baselineskip 0.8cm
\parindent=0.3cm
\parskip=0.3cm
\hspace{0.3cm}
Strangeness production is expected to be a powerful probe for the mechanism
of nucleus-nucleus collisions. Strangeness enhancement, the increased 
strangeness particle production in nucleus-nucleus collisions comparing to 
the nucleon-nucleon collision, is predicted 
to be a sensitive signature of the QGP formation in the ultrarelativistic
nucleus-nucleus collisions [1].

The first experimental results of the enhanced production of strange particles 
in nucleus-nucleus collisions at 200A GeV incident energy was reported
six years ago [2]. Later on this enhancement was confirmed by more and more
experiments [3-7].

Strange antibaryon production, in particular, might bring messages of the 
equilibrium and flavour symmetry of quarks in 
ultrarelativistic nucleus-nucleus collisions.
It has been estimated [8] that if there is QGP formed in the ultrarelativistic 
nucleus-nucleus collisions, strange (antistrange) quark pair might be 
copiously reproduced, resulting in an approximate flavour symmetry among u, 
d and s quarks [9]. Thus the ratio of $\bar{\Lambda}$/$\bar p$ should 
approach unity, since $\bar{\Lambda}$ is composed of $\bar u\bar d\bar s$ and 
$\bar p$ is $\bar u\bar d\bar d$. 

Recently published NA35 data of antibaryon production in sulphur-nucleus 
collisions at 200A GeV [7] did really observe that the ratios $\bar{\Lambda}$ 
/$\bar p$ are all approaching, even over, unity and far exceeding 
the corresponding value of $\sim$0.25 in the nucleon-nucleon collision at the same 
energy. Although the NA35 data of the $\bar{\Lambda}$/$\bar p$ has been fairly 
reproduced by RQMD 1.08 [10,7] with fusion of overlpping strings into a color 
rope and with hadronic rescattering, the data of transverse mass distributions 
of $\bar p$ and $\bar{\Lambda}$ have not been explained yet, the physics behind 
the data have not been exposed especially.

Based on the idea that the strangeness enhancement in ultrarelativistic 
nucleus-nucleus collision compared to the nucleon-nucleon collision at the 
same energy could be investigated via the reduction of s quark suppression 
[11-14], in this letter the event generator LUCIAE [15] is used to 
analyse the NA35 data and to explore the physics behind the data. 
 
LUCIAE was updated based on FRITIOF 7.02 [16] by taking into account the 
collective string interaction [17-18] during the emission of gluon 
bremsstrahlung and the rescattering of produced particles [19]. 
One knows that in FRITIOF 7.02 [16], two colliding hadrons are excited due to
the longitudinal momentum transfer and/or a Rutherford parton scattering. The
highly excited states will emit bremsstrahlung gluons according to the soft
radiation model. The deexcited states are then treated as Lund strings
allowing to decay into the final hadronic state due to the Lund fragmentation 
scheme. However, in the ultrarelativistic nucleus-nucleus collisions there are
generally many excited strings formed close by each other. These strings would
behave like vortex lines in a color superconducting QCD vacuum and will
interact with each other (as the repulsive interaction suffered by the
"ordinary" vortex lines in a type II superconductor). This kind of
collective interaction among strings, not included in FRITIOF 7.02, is
depicted by the firecracker model [15,18] dealing with the large
p$_t$ gluon (firecracker gluon) production from the collective interaction
among strings. The firecracker gluon will work as a gluon-kink excitation on a
string, which tends to shorten the longitudinal size of the string and then
brings about the increasing of the string tension (thus it should be regarded
as an effective string tension). In firecracker model it is  assumed that 
the groups of neighboring strings might form interacting quantum states, the 
large common energy density (corresponding to the collective interaction among
strings) might then affect the emission of gluonic bremsstrahlung [17-18].

A rescattering model has been developed to describe the reinteraction of
produced particles, from FRITIOF event generator, with each other and with
the participant and the spectator nucleons [19]. In the model, the produced
particles and the participant (wounded) nucleons, from FRITIOF event
generator, are distributed randomly in the geometrical overlapping region
between the projectile and the target nuclei under a given impact parameter.
The target (projectile) spectator nucleons are distributed randomly outside the
overlapping region and inside the target (projectile) sphere. A rescattering
cascade process has evolved since then, cf. Ref. [15,19] for the detail.  The
considered inelastic reactions, concerning strangeness, are here cataloged into:
\begin{tabbing}
ttttttttttttttt\=ttttttttttttttt\=tttttt\=tttttttttttttttt\=  \kill
\>$\pi\pi \rightleftharpoons k\bar{k}$;\\
\>$\pi N \rightleftharpoons kY$,
\> \>$\pi\bar{N} \rightleftharpoons  \bar{k}\bar{Y}$;\\
\>$\pi Y  \rightleftharpoons k\Xi$,
\> \>$\pi\bar{Y}  \rightleftharpoons  \bar{k}\bar{\Xi}$;\\
\>$\bar{k}N  \rightleftharpoons  \pi Y$ ,
\> \>$k\bar{N}  \rightleftharpoons  \pi\bar{Y}$;\\
\>$\bar{k}Y  \rightleftharpoons  \pi\Xi$,
\> \>$k\bar{Y}  \rightleftharpoons  \pi\bar{\Xi}$;\\
\>$\bar{k}N  \rightleftharpoons  k\Xi$,
\> \>$k\bar{N}  \rightleftharpoons  \bar{k}\bar{\Xi}$;\\
\>$\pi\Xi \rightleftharpoons k\Omega^- $,
\> \>$\pi\bar{\Xi} \rightleftharpoons  \bar{k}\overline{\Omega^-}$;\\
\>$k\bar{\Xi} \rightleftharpoons \pi\overline{\Omega^-}$,
\> \>$\bar{k}\Xi \rightleftharpoons \pi\Omega^-$;\\
\>$\bar{N}N$ annihilation;\\
\>$\bar{Y}N$ annihilation;\\
\end{tabbing}
where $Y$ refers to the $\Lambda$ or $\Sigma$ and $\Xi$ refers to the $\Xi^-$ 
or $\Xi^0$. There are 299 inelastic
reactions involved altogether. As the reactions introduced above do not make
up the full inelastic cross section, the remainder is again treated as elastic
scattering [19]. The cross section of $\pi\pi \rightarrow k\bar{k}$ is taken
to be 2.0 mb as usual [10]. The isospin averaged parametrization of Ref. [9]
is adopted for the cross sections of the reactions $\pi N \rightarrow kY$ and
for the other strange quark production reactions. Of course, the difference in
threshold energy among reactions is taken into account. Following Ref. [9],
the cross section of strange quark exchange reaction, $\bar{k}N\rightarrow\pi Y
$ for instance, is assumed to be equal to ten times the value of the cross
section of the strangeness production reaction. As for the cross section of
the reverse reaction, the detailed balance assumption [20] is required.

The cross sections of the inelastic reactions given by the isospin averaged
parameterization formulas of [9] for the $\pi N \rightarrow kY$ decrease
exponentially with the CMS energy of the two colliding particles. But we know
that the total inelastic cross section of the $\pi N$ is approximately energy
-independent, which means that more inelastic channels would occur like $\pi^-$
 + p $\rightarrow K^{*0}$ + $\Lambda \rightarrow$ K + $\pi + \Lambda$ and
$\pi^-$ + p $\rightarrow K^{*0}$ + $\Sigma^0 \rightarrow$ K + $\pi + \Sigma^0$
etc., when the CMS energy of the two colliding particles is increasing. In
some sense the reactions we listed above should be looked upon as
'representative channels' of certain types of reactions. Therefore instead of
applying energy-dependent inelastic cross section, we could alternatively give
a constant 'effective cross section' to the reaction $\pi N \rightarrow kY$
[15,19].

In JETSET routine, which runs together with
LUCIAE event generator, there are model 
parameters parj(2) (or 's') and parj(3) which are responsible for the s quark 
suppression and related to the effective string tension. 's' refers to the 
suppression of s quark pair production in the color field compared to u or d 
pair production. parj(3) is the extra suppression of strange diquark 
production compared to the normal suppression of strange quark pair. Besides 
's' and parj(3) there is parj(1), which stands for the suppression of diquark-
antidiquark pair production in the color field compared to the quark-antiquark 
pair production and is related to the effective string tension as well. 
Originally, in the LUND fragmentation scheme 's' was assumed to be a 
'constant' and this assumption was confirmed by the e$^+$e$^-$ physics from 
the low energies to Z$^0$ energy. However, there are experimental facts that 
this parameter is energy dependent when the fragmentation scheme is applied to 
the Deep Inelastic Scattering (DIS) and hh collisions. In DIS experiments, 's' 
is seen to be rising up with the increase of energy from 0.15 at $\sqrt{S}$=5 
GeV to 0.35 at $\sqrt{S}$=20 GeV [21].
An energy dependent 's' for hh collisions has been known for
many years. It varies from 0.2 at the ISR energy to about 0.4 at $\sqrt{S}$=
1.8 TeV [21]. In addition, this parameter is also observed strongly phase
space dependence in DIS and hh collisions. The corresponding value of 's'
runs from 0.15 to 0.55 with a mean value close to 0.3 [22]. Thus the idea of  
the reduction of s quark suppression in nucleus-nucleus collisions comparing 
to the nucleon-nucleon collision is possible to be executed via changing the 's' 
and concerning parameters in JETSET routine.

From LUND string model point of view s quark suppression factor  
could be related to the effective string tension. It is reasonable to expected 
that the more violent collision, the stronger collective interaction among 
strings, the larger effective string tension and then the larger s
quark suppression factor.  Generally speaking, a nucleus-nucleus collision is 
more violent than the nucleon-nucleon collision at the same interaction energy. 
That might be the reason of the reduction of s quark suppression in nucleus-
nucleus collisions comparing to the nucleon-nucleon collision at the same energy.

The purpose of this letter is to explore the physics behind the NA35 data and  
not to fit the data as good as possible. We fix a set of somewhat larger 
parameters (relative to the defaults: 's'=0.3, parj(3)=0.4 and parj(1)=0.1) of 
's'=0.4, parj(3)=0.5333 and parj(1)=0.1333 (referred to as parameter set 1, 
later on) in the calculations of nucleus-nucleus collisions and a set of 
somewhat smaller parameters of 's'=0.2, parj(3)=0.2666 and parj(1)=0.06666 (
referred to as parameter set 2) in the calculation of the 
nucleon-nucleon collision 
. These results are given in table 1 and figure 1 and obtained from the 
average over 10$^5$ generated events for pp, 2000 events for $\bar{\Lambda}$ 
and $\bar{p}$ yield in AA, and 3000 for m$_t$ distribution in AA. The y and 
p$_t$ acceptances are set to be the same as the experiment [7], correspondingly. 
Of course, the two sets of parameters above imply that one makes an assumption 
there that those three parameters are linearly proportional to the effective 
string tension in their responsibilities to the final hadronic production.

Tab. 1 gives the rapidity densities of $\bar{p}$, $\bar{\Lambda}$, and $h^-$
and the ratio $\bar{\Lambda}$/$\bar{p}$
in p + p and S + S, S + Ag, and S + Au central collisions at 200A GeV. The 
rapidity acceptance is 3 $\leq$ y  $\leq$ 4. The results of LUCIAE are 
comparable with the corresponding NA35 data. The LUCIAE results of $\bar{\Lambda}$
/$\bar{p}$ seem going down monotonously from S + S to S + Au, which might 
attribute to the fact that one did not consider the possible difference of 
the reduction of strangeness suppression among them. The results of LUCIAE 
and RQMD are comparable with each other, since both of them do have contained  
the collective interaction among strings and the hadronic
rescattering [23]. The hadronic rescattering, in LUCIAE and RQMD, is expected
to play a similar role in the final state distributions though details of the
hadronic rescattering are not the same in these two generators. In
RQMD the effect of highly dense strings in ultrarelativistic nucleus-nucleus
collisions is considered via the fusion of strings into a color rope, which
brings the enhanced production of strange antibaryon.
In LUCIAE the highly dense strings are considered by the firecracker model
[15,18], where the effect of highly dense strings is depicted
analytically only in the emission of gluon bremsstrahlung. It leaves the
possibility to enlarge the s quark pair suppression factor and concerns in 
JETSET routine to bring the enhanced production of strange antibaryon. 

Fig. 1 gives $\bar{p}$ and $\bar{\Lambda}$ transverse mass distributions  
in central S + S (upper frame), S + Ag (middle frame), and S + Au (lower 
frame) reactions at 200A GeV, respectively.   
In this figure, the open squares and open circles are the NA35 
data of $\bar{p}$ and $\bar{\Lambda}$, respectively and the corresponding 
results of LUCIAE are given by full squares and full circles. The solid lines 
are the exponential fits of the NA35 data, cf. Ref. 7 for details. One sees 
from this figure that the agreement between the NA35 data and the results of 
LUCIAE is fair for $\bar{p}$ and reasonably good for $\bar{\Lambda}$.

Table 2 gives the average yield (in full phase space) of $\Lambda$, 
$\bar{\Lambda}$, K$^+$ and K$^0_s$, in p+p and central S + S and S + Ag 
collisions at 200A GeV. In this table the data were taken from [5] besides p+p,
which was taken from [24]. One sees from this table that although we are here 
aiming at analysing strange antibaryon production data, the strange particle 
production data are also reproduced reasonably good at the same time.    

In order to distinguish the role of the s quark pair suppression factor and 
concerns from the hadronic rescattering and the role of 's' and parj(3) from 
parj(1) one calculates the results of table 3 and figure 2. In table 3 one 
compares the results of the rapidity densities of $\bar{p}$, 
$\bar{\Lambda}$, and $h^-$ and the ratio $\bar{\Lambda}$/$\bar{p}$ in  S + S 
central collisions at 200A GeV calculated by using different parameters and 
with or without rescattering: 'LUCIAE 1': parameter set 2 and without 
rescattering; 'LUCIAE 2': parameter set 2 and with rescattering; 'LUCIAE 3': 
parameter set 1 but the value of parj(1) is changed to 0.06666 and with 
rescattering; and 'LUCIAE 4': parameter set 1 and with rescattering. One sees 
from this table that the 'LUCIAE 1' result of $\bar{\Lambda}$/$\bar{p}$ is 
close to the corresponding result in the pp collision, as it should be. For 
results of 'LUCIAE 2', $\bar{\Lambda}$ yield is close to the corresponding 
result of 'LUCIAE 1' but $\bar{p}$ yield is lower than 'LUCIAE 1', which brings 
about the larger ratio $\bar{\Lambda}$/$\bar{p}$ in 'LUCIAE 2' than in 'LUCIAE 
1'. Here one knows that the hadronic rescattering seems to play nearly null 
role for ${\bar\Lambda}$ yield, since ${\bar\Lambda}$ production in 
rescattering is mainly via $\pi\bar{N} \rightleftharpoons\bar{k}\bar{Y}$ and 
$k\bar{N} \rightleftharpoons \pi\bar{Y}$, that is nearly canceled by the 
corresponding inverse reactions and $\bar{Y}N$ annihilations especially. 
On the contrary, the rescattering tends to reduce $\bar{p}$ multiplicity 
through the $\bar{p}p$ annihilation. However, relying on hadronic rescattering 
only is not possible to have the ratio $\bar{\Lambda}$/$\bar{p}$ approaching 
to unity. Furthermore, comparing the results of 'LUCIAE 3' with 'LUCIAE 4' 
one knows that 
although by increasing  parj(1) the yield of $\bar{p}$ and $\bar{\Lambda}$ are 
increased the ratio $\bar{\Lambda}$/$\bar{p}$ is hardly affected.

Figure 2 gives the transverse mass 
distributions of ${\bar\Lambda}$ (upper frame) and $\bar{p}$ (lower frame) in 
S + S reaction at 200A GeV, respectively. The full triangles,  
the open squares, and the full circles are calculated individually  
for the case 1: parameter set 2 and with rescattering; the case 2: parameter 
set 1 and without rescattering; and the case 3: parameter set 1 and with 
rescattering. The rapidity acceptance are all set to be 
1 $\leq$ y  $\leq$ 3. One knows from this figure again that the hadronic 
rescattering seems to play nearly null role for ${\bar\Lambda}$ production,  
but hadronic rescattering tends to reduce $\bar{p}$ multiplicity, at lower 
m$_t$ region especially, through the $\bar{p}p$ annihilation. 

In summary, we have roughly reproduced the NA35 data of $\bar p$, 
$\bar{\Lambda}$ yields , the corresponding ratio $\bar{\Lambda}$/$\bar p$, and  
the transverse mass distributions of $\bar p$ and $\bar{\Lambda}$ in pp and 
central S + S, S + Ag, and S + Au collisions at 200A GeV, using event 
generator LUCIAE and via increasing s quark pair suppression factor and concerns 
in AA collisions. It seems to be true that the NA35 data [5,7] imply the 
reduction of strangeness suppression in ultrarelativistic nucleus-nucleus 
collisions comparing to the pp collision at the same energy. Although the ratio 
$\bar{\Lambda}$/$\bar p$ approaching to unity 
is the result of reduction of s quark suppression, it does not necessarily 
mean a flavor symmetry, since reproducing the NA35 data does not require 
the 's' value should be equal to one and the hadronic rescattering also 
plays a role in enlarging the ratio $\bar{\Lambda}$/$\bar p$.
However, it is absolutely needed to have a further study for 
the microscopic mechanism of the reduction of strangeness suppression before 
making a conclusion using strangeness enhancement as a signal of QGP.

\begin{center}Acknowledgment\end{center}
We are grateful to J. Eschke and D. R$\ddot{o}$hrich for providing the NA35 
data. Thanks go to O. Miyamura, K. Kumagai and T. Sasaki for discussions and 
helps. SBH thanks Department of Physics, University of Hiroshima for 
hospitality and JSPS for financial support staying in Japan to finish most 
of the calculations. This work is supported by the national Natural Science 
Foundation of China as a cooperation program between NSFC of China and JSPS of 
Japan.

\begin{center}References\end{center}
\begin{enumerate}
\item M. Jacob and J. Tran Van., Phys. Rep., $\bf{88}$, 321(1982);\\
      J. Rafelski, Phys. Rep., $\bf{88}$, 331(1982);\\
      P. Koch and J. Rafelski, Nucl. Phys., $\bf{A444}$, 678(1985);\\
      J. Ellis and U. Heinz, Phys. Lett., $\bf{B233}$, 223(1989).
\item J. Bartke, et al., NA35 Colla., Z. Phys., $\bf{C48}$, 191(1990).
\item E. Andersen, et al., NA36 Colla., Phys. Lett., $\bf{B316}$, 603(1993).
\item S. Abatzis, et al., WA85 Colla., Phys. Lett., $\bf{B244}$, 127(1990).
\item T. Alber, et al., NA35 Colla., Z. Phys., $\bf{C64}$, 195(1994).
\item E. Andersen, et al., NA36 Colla., Nucl. Phys., $\bf{A590}$, 291c(1995).
\item T. Alber, et al., NA35 Colla., Phys. Lett., $\bf{B366}$, 56(1996).
\item E. V. Shuryak, Phys. Rep.,  $\bf{61}$, 71(1980); $\bf{115}$, 151(1984).
\item P. Koch, B. M$\ddot{u}$ller, and J. Rafelski, Phys. Rep., $\bf{142}$, 167
       (1986).
\item H. Sorge, Phys. Rev., $\bf{C52}$, 3291(1995).
\item Sa Ben-Hao and Tai An, Phys. Rev., $\bf{C55}$, 2010 (1997)
\item A. K. Wr$\acute{o}$blewski, Acta Phys. Pol., $\bf{B16}$, 379(1985).
\item H. Bialkowska, M. Ga$\acute{z}$dzicki, W. Retyk, and E. Skrzypczak, 
       Z. Phys., $\bf{C55}$, 491(1992).
\item M. Ga$\acute{z}$dzicki, and U. Heinz, Phys. Rev., $\bf{C54}$, 1496(1996).
\item Sa Ben-Hao and Tai An, Comp. Phys. Commu., $\bf{90}$, 121(1995).
\item B. Andersson, G. Gustafson, and Hong Pi, Z. Phys., $\bf{C57}$, 485(1993).
\item B. Andersson, Phys. Lett., $\bf{B256}$, 337(1991).
\item B. Andersson and An Tai, Z. Phys., $\bf{C71}$, 155(1996). 
\item Sa Ben-Hao, Wang Zhong-Qi, Zhang Xiao-Ze, Song Guang, Lu Zhong-Dao,
       and Zheng Yu-Ming, Phys. Rev., $\bf{C48}$, 2995(1993);\\
      Sa Ben-Hao, Tai An, and Lu Zhong-Dao, Phys. Rev., $\bf{C52}$, 
      2069(1995);\\
      B. Andersson, An Tai and Ben-Hao Sa, Z. Phys., $\bf{C70}$, 499(1996).
\item G. Bertsch, and S. Das Gupta, Phys. Rep., $\bf{160}$, 189(1988).
\item A. K. Wr$\acute{o}$blewski, Proceedings of the 25th International 
conference on HEP, p. 125, Singapore, 1990.
\item ZEUS Colla., Z. Phys., $\bf{C68}$, 29(1995);\\
      Fermilab E665 Colla., Z. Phys., $\bf{C61}$, 539(1994).
\item H. Sorge, Z. Phys., $\bf{C67}$, 479(1995).
\item M. Ga$\acute{z}$dzicki, and O. Hansen, Nucl. Phys., $\bf{A528}$, 754
      (1991).
\end{enumerate}
\newpage
\begin{center}Figure Captions\end{center}
\begin{quotation}
Fig. 1 Transverse mass distributions of $\bar p$ (3 $\leq$ y $\leq$ 4) and 
$\bar\Lambda$ produced in central S+S (upper frame, 1 $\leq$ y $\leq$ 3), 
S+Ag (middle frame, 1 $\leq$ y $\leq$ 3), and S+Au (lower frame, 3  $\leq$ 
y $\leq$ 5) collisions at 200A GeV. The open squares and open circles  
are the NA35 data of $\bar p$ and $\bar\Lambda$, 
respectively and the full squares and full circles are the corresponding 
results of LUCIAE.

Fig. 2 Transverse mass distributions (1 $\leq$ y $\leq$ 3) of $\bar\Lambda$ 
(upper frame) and $\bar p$ (lower frame) in  
S+S reaction at 200A GeV. The full triangles, the open squares, and the full  
circles are calculated individually for the case 1: parameter set 2 ('s'=0.2, 
parj(3)=0.2666 and parj(1)=0.06666) and with rescattering; the case 2:  
parameter set 1 ('s'=0.4, parj(3)=0.5333, parj(1)=0.1333) and without 
rescattering; and the case 3: parameter set 1 and with rescattering.
\end{quotation}

\newpage
\vspace{0.5cm}
\begin{tabular}{cccccc}
\multicolumn{6}{c}{Table 1. Rapidity densities (3 $\leq$ y $\leq$ 4) of 
		   $\bar p$, $\bar\Lambda$ and h$^-$} \\
\multicolumn{6}{c}{produced in p+p and 
		   central sulphur-nucleus collisions at 200A GeV}\\  
\hline\hline
 reaction & &$\bar p$ & $\bar\Lambda$ &  $\bar\Lambda$/$\bar p$ 
	  &h$^-$ \\
\hline
        p + p& data & 0.02$\pm$0.02  & 0.005$\pm$0.002    &
                          0.25$\pm$0.1    & 0.74$\pm$0.04 \\
         &LUCIAE     & 0.017  & 0.0035    & 0.21    &  0.60\\
\bigskip
             &RQMD & 0.015  &0.005     &  0.3   &  $-$\\ 
        S + S& data & 0.4$\pm$0.1  & 0.76$\pm$0.16    &
                          1.9$^{+0.7}_{-0.6}$ &25$\pm$1\\ 
central &LUCIAE     & 0.65  & 0.66    & 1.03    & 23.7 \\
\bigskip
             &RQMD & 0.7  & 0.75   & 1.1    & $-$ \\ 
       S + Ag& data & 0.6$\pm$0.2  & 0.75$\pm$0.19    &
                          1.3$^{+0.7}_{-0.5}$    & 40$\pm$2 \\
central&LUCIAE     & 1.08  & 0.99    & 0.91    & 39.3 \\
\bigskip
            &RQMD & 1.0  & 0.9    & 0.9   & $-$ \\ 
       S + Au& data & 0.7$\pm$0.2  & 0.75$\pm$0.1   &
  1.1$^{+0.4}_{-0.3}$    & 47$\pm$5 \\ 
central&LUCIAE     & 1.09  & 0.91    & 0.84    & 42.7 \\
	    &RQMD & 1.4  & 1.2    & 0.9   & $-$ \\
\hline
\hline
\end{tabular}

\vspace{1.5cm}
\begin{tabular}{cccccc}
\multicolumn{6}{c}{Table 2. Average yield (in full phase space) of $\Lambda$, 
                   $\bar{\Lambda}$, K$^+$ and K$^0_s$}\\ 
\multicolumn{6}{c}{produced in p+p and
                   central sulphur-nucleus collisions at 200A GeV}\\
\hline\hline
 reaction & & $\Lambda$ & $\bar\Lambda$ & K$^+$ & K$^0_s$ \\
\hline
p + p& data & 0.096$\pm$0.015 & 0.013$\pm$0.005 & $-$ & 0.17$\pm$0.01 \\
\bigskip
& LUCIAE & 0.10 & 0.011 & 0.22 & 0.16 \\
S + S& data & 9.4$\pm$1.0 & 2.2$\pm$0.4 & $-$ & 10.5$\pm$1.7 \\
\bigskip
central & LUCIAE & 8.0 & 1.9 & 11.9 & 9.83 \\
S + Ag& data & 15.2$\pm$1.2 & 2.6$\pm$0.3 & $-$ & 15.5$\pm$1.5 \\
central & LUCIAE & 13.9 & 3.1 & 21.3 & 17.6 \\
\hline
\hline
\end{tabular}

\vspace{1.5cm}
\begin{tabular}{cccccc}
\multicolumn{6}{c}{Table 3. Rapidity densities (3 $\leq$ y $\leq$ 4) of
		   $\bar p$, $\bar\Lambda$ and h$^-$} \\
		   \multicolumn{6}{c}{produced in p+p and center S+S collision
		   at 200A GeV}\\
\hline
 reaction & &$\bar p$ & $\bar\Lambda$ &  $\bar\Lambda$/$\bar p$
	   &h$^-$ \\
\hline
	p + p& data & 0.02$\pm$0.02  & 0.005$\pm$0.002    &
			  0.25$\pm$0.1    & 0.74$\pm$0.04 \\
   &LUCIAE     & 0.017  & 0.0035    & 0.21    &  0.60\\
\bigskip
	&RQMD & 0.015  &0.005     &  0.3   &  $-$\\
	S + S& data & 0.4$\pm$0.1  & 0.76$\pm$0.16    &
			  1.9$^{+0.7}_{-0.6}$ &25$\pm$1\\
central &LUCIAE 1$^a$     & 0.62  & 0.17    & 0.28    & 25.2 \\
&LUCIAE 2$^b$    & 0.42  & 0.17    & 0.40    & 24.8 \\
&LUCIAE 3$^c$    & 0.36  & 0.38    & 1.05    & 24.7 \\
&LUCIAE 4$^d$    & 0.65  & 0.66    & 1.03    & 23.7 \\
\hline
\hline
\multicolumn{6}{c}{a. 's'=0.2, parj(3)=0.2666, parj(1)=0.06666 without 
		   rescattering}\\
\multicolumn{6}{c}{b. 's'=0.2, parj(3)=0.2666, parj(1)=0.06666 with
		   rescattering}\\
\multicolumn{6}{c}{c. 's'=0.4, parj(3)=0.5333, parj(1)=0.06666 with
		   rescattering}\\
\multicolumn{6}{c}{d. 's'=0.4, parj(3)=0.5333, parj(1)=0.1333 with
		   rescattering}\\
\end{tabular}

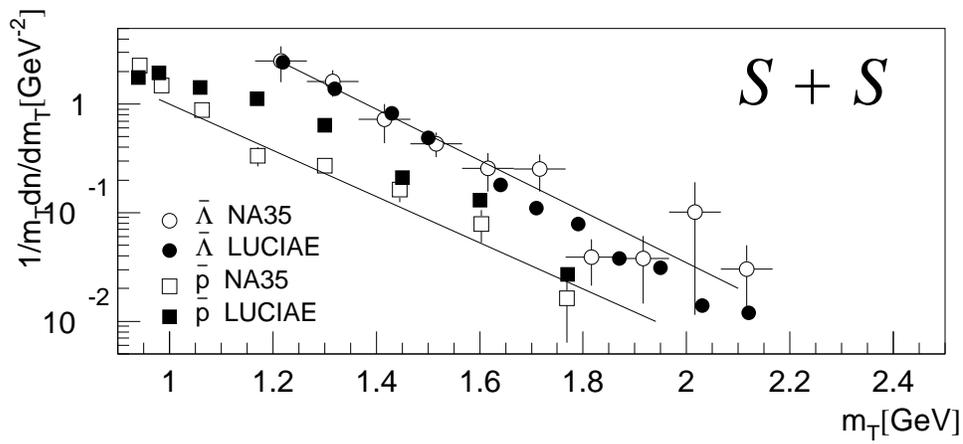
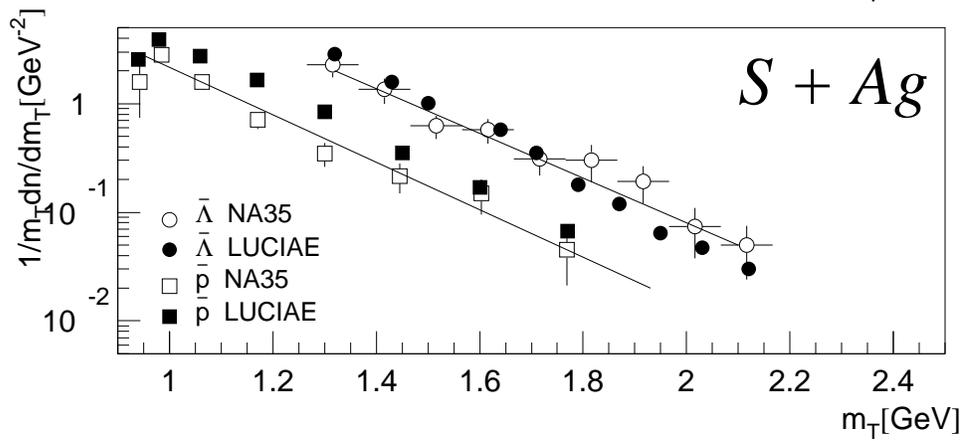
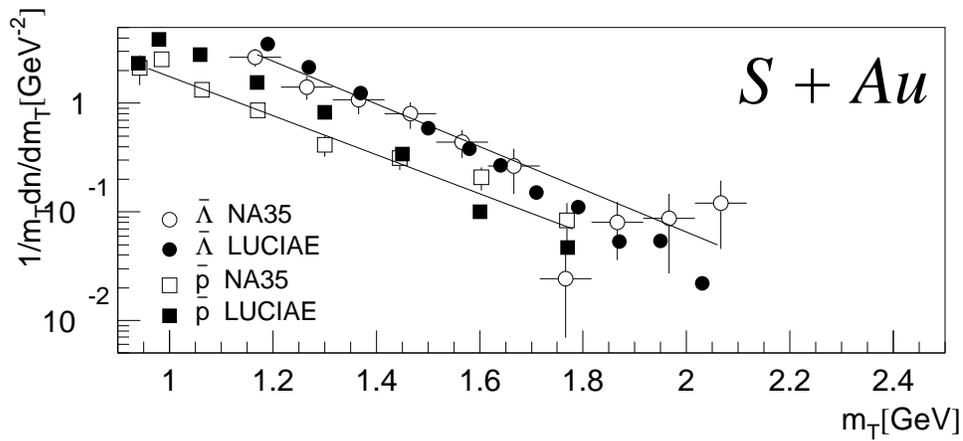
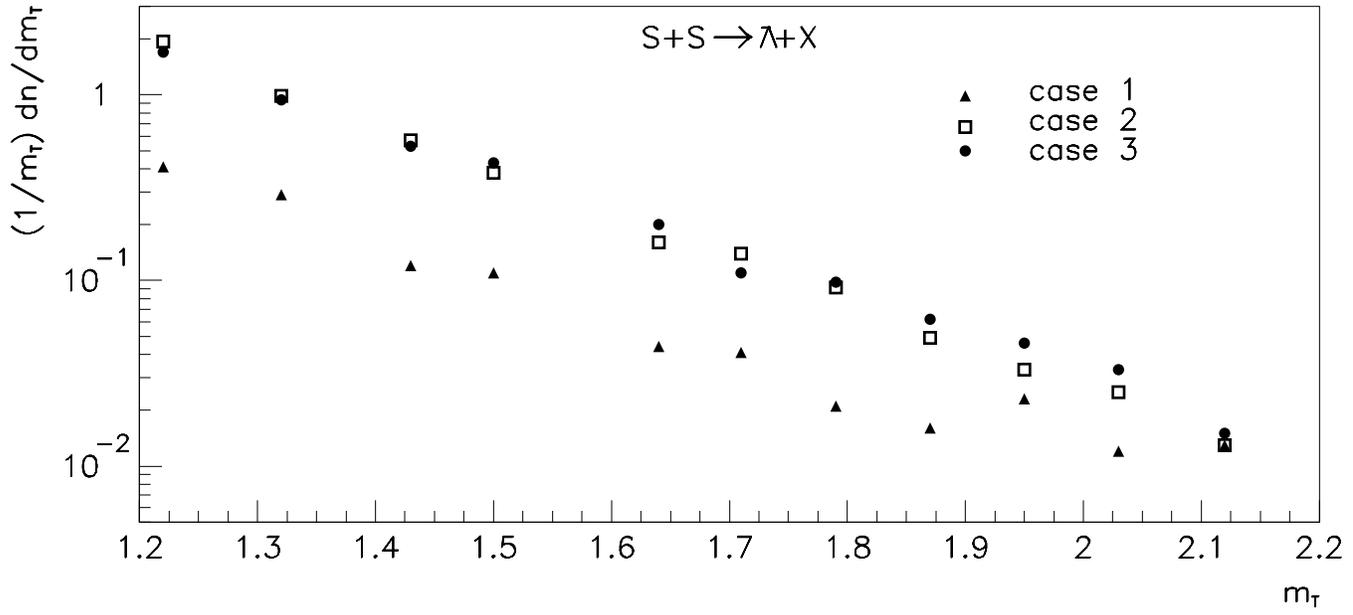
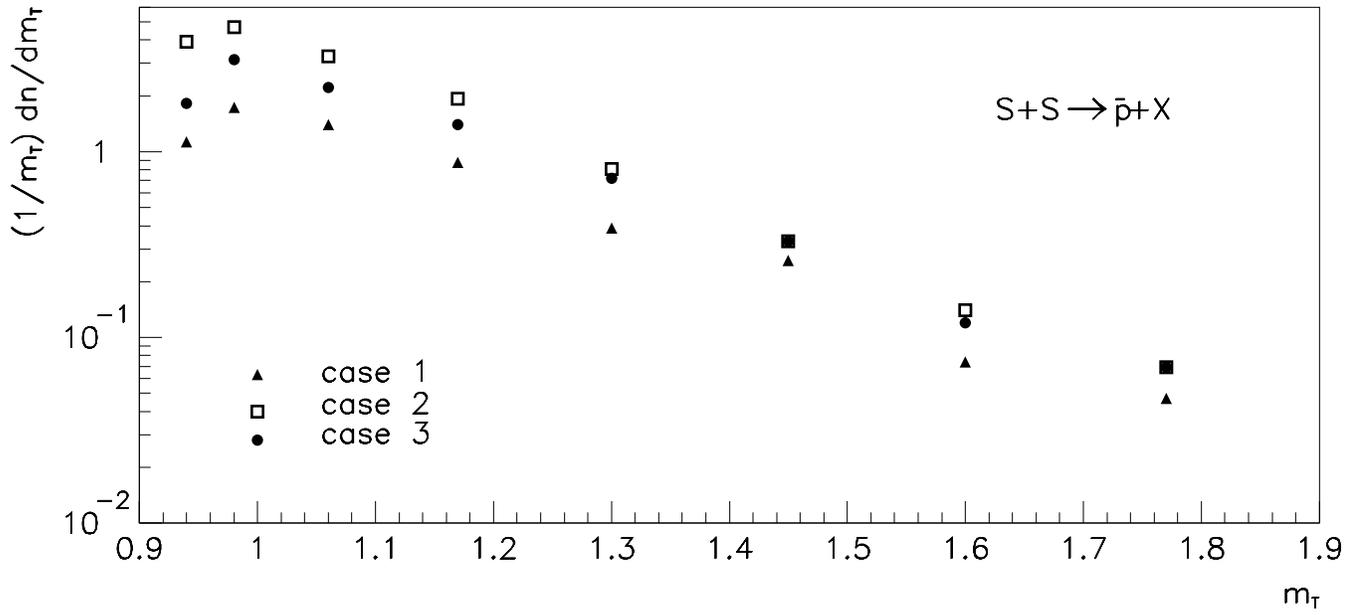
\end{document}